\documentstyle[11pt,aaspp4,psfig]{article}

\def\dash{\hbox{--}}

\begin{document}

\title{Attenuation of Beaming Oscillations Near Neutron Stars}

\author{M. Coleman Miller}
\affil{Department of Astronomy, University of Maryland\\
       College Park, MD  20742-2421\\
       miller@astro.umd.edu}
\authoremail{miller@astro.umd.edu}

\begin{abstract}

Observations with the {\it Rossi} X-ray Timing Explorer have
revealed kilohertz quasi-periodic brightness oscillations
(QPOs) from nearly twenty different neutron-star low-mass X-ray binaries
(LMXBs).  These frequencies often appear as a pair of kilohertz QPOs
in a given power density spectrum.  It is extremely 
likely that the frequency of the
higher-frequency of these QPOs is the orbital frequency of gas
at some radius near the neutron star. It is also likely that the QPOs
are caused by the movement of bright arcs or luminous clumps around
the star, which produce a modulation in the observed X-ray intensity 
as they are periodically occulted by the star or as they present a 
different viewing aspect to the observer at infinity. If this
picture is correct, it means that this type of QPO is a beaming
oscillation. In such models it is expected that there will also be 
beaming oscillations at the stellar spin frequency and at
overtones of the orbital frequency, but no strong QPOs have
been detected at these frequencies.

We therefore examine the processes that can  attenuate beaming
oscillations near neutron stars, and in doing so extend the work
on this subject that was initiated by the discovery of
lower-frequency QPOs from LMXBs. We consider attenuation by
scattering, attenuation by light deflection, and the decrease in
modulation caused by integration over the visible  surface of the
neutron star. Our main results are (1)~in a spherical scattering
cloud, {\it all} overtones of rotationally modulated beaming
oscillations are attenuated strongly, not just the even
harmonics,  (2)~the amount of attenuation is diminished, and hence
the observed modulation amplitude is increased, by the presence of
a central, finite-sized star, even if the scattering cloud is much
larger than the star, and (3)~if the specific intensity of
radiating points on the star has a large angular width, then even
with zero optical depth from the stellar surface to the observer,
and even in the approximation of  straight-line photon
propagation, the modulation amplitude seen at infinity is
decreased significantly by integration over the visible portion of
the surface.  We also compare the  modulation of flux as seen at
infinity with the modulation  near the star, and show that (4)~it
is possible to have a relatively high-amplitude modulation near
the star at, e.g., the stellar spin  frequency, even if no peak at
that frequency is detectable in a power  density spectrum taken at
infinity.

\end{abstract}

\section{INTRODUCTION}

The discovery of kilohertz quasi-periodic brightness oscillations
(QPOs) from many neutron-star low-mass X-ray binaries (LMXBs) has
given us a sensitive probe of the conditions near accreting neutron stars
(see van der Klis 1998 for a review of the properties of kilohertz QPOs).
A variety of models have been suggested for this phenomenon,
including beat-frequency models (Strohmayer et al.\ 1996; Miller,
Lamb, \& Psaltis 1998), relativistic precession models (Stella \& Vietri 1998),
and disk oscillation models (e.g., Osherovich \& Titarchuk 1999).

In any such model, the brightness oscillations can be produced in two
general ways. In a pure {\it luminosity} oscillation, the 
total luminosity of the source changes quasi-periodically but the
angular distribution of the radiation remains fixed. In a pure
{\it beaming} oscillation, the total luminosity remains
constant but the angular distribution of the radiation changes and
hence the flux in the direction of the observer is
modulated. An example of a luminosity
oscillation is the beat-frequency oscillation in the magnetospheric
beat-frequency
model proposed for the so-called horizontal branch oscillation
seen from Z sources (Alpar \& Shaham 1985; Lamb et al. 1985; Shibazaki
\& Lamb 1987). An example of
a beaming oscillation is the modulation created by an accretion-powered
pulsar, in which, as the neutron star rotates, the hot spots on 
the magnetic polar cap pass into and out of our view, creating an
observed modulation.

Here we focus on beaming oscillations. In some models of the kilohertz
QPOs (e.g., Miller et al.\ 1998), the QPO peak with the highest
frequency in a given source is a beaming
oscillation caused by the movement of bright arcs of impact around
the star, at the orbital frequencies in a special range of radii in the 
accretion disk. Only the fundamental for such an oscillation has
been observed, and the amplitudes at any overtones are down by
at least factors of several. Moreover, in no kilohertz QPO source has there 
been a strong brightness oscillation observed at the putative stellar spin
frequency during the persistent emission between type 1 X-ray
bursts (kilohertz QPOs have not been observed from the millisecond X-ray
pulsar SAX~J1808-3658; see Wijnands \& van der Klis 1998).  For some of the 
higher-luminosity
sources, such as Sco~X-1, the upper limit on the pulsed fraction is
as low as 0.3\% (Vaughan et al.\ 1994).  This constrains beat frequency
models, in which the 
lower-frequency QPO peak is created in part by a modulation at the
spin frequency. It is therefore necessary to explain how the spin
modulation can be important close to the star but undetectable by
current instruments at infinity.

To understand this, we need to consider various ways in which beaming
oscillations can be attenuated on their way to us. Two general
effects are (1)~integration over the visible surface of
the star, which tends to smooth out variations in intensity, and
(2)~scattering in a surrounding hot central corona, whose existence
has been inferred from previously existing spectral and temporal
modeling of neutron-star LMXBs (Lamb 1989; Miller \& Lamb 1992;
Psaltis, Lamb, \& Miller 1995;
Psaltis \& Lamb 1997). Previous treatments of the 
attenuation of beaming oscillations by scattering (Brainerd \&
Lamb 1987; Kylafis \& Phinney 1989) assumed for simplicity that
the source of radiation was a pencil beam at the center of
a scattering cloud with uniform electron density (although
see Wang \& Schlickeiser 1987 for a discussion of offset
emission and varying electron density in the diffusion
approximation), and have implicitly assumed
that photons follow straight-line trajectories (although for a
treatment of general relativistic light-bending without scattering,
see Wood, Ftaclas, \& Kearney 1988; M\'esz\'aros, Riffert, \& 
Berthiaume 1988). With the large amounts
of data available from the {\it Rossi} X-ray Timing Explorer (RXTE), 
it is now important to examine some
of the deviations expected from this idealization.

Here we calculate the attenuation of beaming oscillations near 
neutron stars, taking into account the finite size of the
star, the varying electron number density near
the star, and general relativistic light deflection. In \S~2
we explain our numerical method. In \S~3 we discuss the decrease
in the modulation amplitude
of beaming oscillations caused by integration over the visible surface of
the star. In \S~4 we
treat several aspects of the attenuation of beaming oscillations
by scattering, including the presence of a finite object, varying
electron number density, modulation from multiple harmonics of
emission, the angular dependence of the specific intensity, and
general relativistic light deflection. In \S~5 we calculate the
modulation amplitude close to the star, within the scattering cloud.
In \S~6 we summarize and give our conclusions.

\section{NUMERICAL METHOD}

The numerical results in this paper are generated using a Monte
Carlo code. We assume that a star of radius
$R$ is in the center of a scattering cloud.
We also define optical depths $\tau_R$ and $\tau_c$ 
such that if the locally measured 
electron number density were uniform then, if the star were not
present, the optical depth from the center of the
sphere to the radius of the star would be $\tau_R$, and the optical
depth from the center to the outer boundary
of the scattering sphere would be $\tau_c$. Photons are
introduced into the sphere at some radius and angle, traveling
in some given direction, and they are followed individually as
they scatter in the sphere.
When the photons escape, their direction is
stored. After all the photons ($10^4$ to $10^6$, depending on the
run) have been tracked, rms amplitudes or beaming ratios can be
computed.

We now discuss how photon paths between scatterings and after escape
are followed for straight-line propagation and light deflection.

\subsection{Straight-Line Photon Propagation}

The number $x$ of mean free paths traversed between two
scatterings is selected randomly according to $e^{-x}$. If
the scattering cloud has a uniform density then the physical
distance traveled is proportional to $x$, otherwise the
distance must be calculated from the density distribution.
Once the distance between two given scatterings is ascertained,
the code checks whether (1)~the photon
hits the star, or (2)~the photon escapes. If the photon escapes,
the current direction of propagation of the photon is stored.
If the photon hits the star, then in the next iteration the
photon is assumed to be emitted in an outward direction
from the point of impact, selected from an isotropic distribution. 
Otherwise, the new location
is calculated by adding the vector of the photon path to the
previous position.

\subsection{Light Deflection}

When light deflection is included, the propagation of photons
is slightly more complicated than it is if photons propagate
in straight lines. Although light deflection occurs in
Newtonian gravity as well as in general relativity, light
deflection is only important when gravity is strong enough
to make general relativistic corrections to Newtonian gravity
significant. Hence, we consider
an external spacetime that is the Schwarzschild spacetime, which
is appropriate outside spherically symmetric, nonrotating
neutron stars. This spacetime includes most of the important
features of, e.g., light deflection around neutron stars, and
the results are easier to evaluate than the results of scattering
around rotating neutron stars. 

In the Schwarzschild spacetime, any geodesic can be treated as an
equatorial geodesic by an appropriate rotation of the coordinate
system (a statement that is not true for general spacetimes with
rotation). This feature means that a simple way to treat photon
propagation between scatters is to follow the path in that
temporary ``equatorial plane" and then rotate back into the global
coordinate system. Hence, there are three new tasks brought up
by the inclusion of light deflection: (1)~calculation of temporary
equatorial planes, (2)~computation of the curved trajectory
of the photon in that plane, and (3)~determination of the 
additional curvature of the photon trajectory after it escapes
from the scattering cloud. We now treat these in order.

Let the current angular location of the photon have a colatitude
$\theta$ and an azimuthal angle $\phi$ in the global coordinate
system. We represent the unit vector in this direction by
$(\theta,\phi)$, which in Cartesian coordinates is as usual the
three-vector $(\sin\theta\cos\phi,\sin\theta\sin\phi,\cos\theta)$,
where the first, second, and third components are along the global
$x$, $y$, and $z$ directions, respectively.
Let the initial direction of propagation of the
photon, also in the global coordinate system, be $(\alpha,\beta)$. 
We set up the new equatorial
plane as follows. The ${\hat x}$-axis is in the direction
$(\theta,\phi)$, and since this is properly normalized we assign
${\hat x}=(\theta,\phi)$. The ${\hat z}$ axis is perpendicular to
the plane containing $(\theta,\phi)$ and $(\alpha,\beta)$, and
is thus in the direction $(\theta,\phi)\times(\alpha,\beta)$. The
remaining axis is in the direction ${\hat y}={\hat z}\times{\hat x}$.
If the angle between $(\theta,\phi)$ and $(\alpha,\beta)$ is
$\psi$, then the unit vectors are
\begin{equation}
\begin{array}{l}
{\hat x}=(\theta,\phi)\\
{\hat y}={1\over{\sin\psi}}\left[(\alpha,\beta)-(\theta,\phi)
\cos\psi\right]\\
{\hat z}={1\over{\sin\psi}}(\theta,\phi)\times(\alpha,\beta)\; ,\\
\end{array}
\end{equation}
and the equatorial plane is defined by ${\hat x}$ and ${\hat y}$.

Thus, if we use $\phi^\prime$ to denote the azimuthal angle in
this new equatorial plane [where $\phi^\prime\equiv 0$ at the original
angular location $(\theta,\phi)$], then the location in the original,
global Cartesian
coordinates at an arbitrary radius $r$ and angle $\phi^\prime$ is
just $r\cos\phi^\prime{\hat x}+r\sin\phi^\prime{\hat y}$.

The next task is to follow the propagation of a photon in this
new equatorial plane between scatterings. To do this, we note that
the locally measured spacelike components of the photon
four-velocity are just
\begin{equation}
\begin{array}{l}
u^{\hat r}=\cos\psi\\
u^{\hat\phi}=\sin\psi\; .\\
\end{array}
\end{equation}
Here we use geometrized units in
which $c=G\equiv 1$. In this and subsequent equations,
hatted quantities such as $u^{\hat r}$ are
measured in a local tetrad, in contrast to unhatted quantities
such as $u^r$, which are measured in the global Boyer-Lindquist
coordinate system. The components of the four-velocity in global
Boyer-Lindquist coordinates are given by the transformation
of these quantities from the local to global frames (see,
e.g., Abramowicz, Ellis, \& Lanza 1993; Miller \& Lamb
1996): 
\begin{equation}
\begin{array}{l}
u^r=(1-2M/r)^{1/2}u^{\hat r}\;\; {\rm and}\\
{u^\phi}=u^{\hat\phi}/r\; .\\
\end{array}
\end{equation}
The photon path is followed by moving a small distance $ds$ along
the ray (we found $ds=0.02$ in units of the mean free path
gives sufficient accuracy) and recalculating the propagation 
angle $\psi$:
\begin{equation}
\sin\psi_{\rm new}=\sin\psi_{\rm old}\left(r_{\rm old}\over{
r_{\rm new}}\right){(1-2M/r_{\rm new})^{1/2}\over{
(1-2M/r_{\rm old})^{1/2}}}
\end{equation}
(see, e.g., Abramowicz et al. 1993 or Miller \& Lamb 1996).
Using this formula,
we can determine the total deflection $\phi^\prime$ and 
rotate back from the temporary equatorial plane to the
global coordinate system.

The last task is to follow the deflection of the photon after
it has escaped. This is done straightforwardly using the approach
of, e.g., Pechenick, Ftaclas, \& Cohen (1983). In this approach,
we define the impact parameter $b=(\sin\psi)r(1-2M/r)^{-1/2}$,
and let $u_b=M/b$. The total deflection angle from radius $r$ to
infinity is (Pechenick et al. 1983, eq. [2.12])
\begin{equation}
\Delta\phi=\int_0^{M/r}\left[u_b^2-(1-2u)u^2\right]^{-1/2}du\; .
\end{equation}
\noindent Note that, as defined, this is actually the difference between the
global azimuthal angle at infinity $\phi(\infty)$ and the global
azimuthal angle $\phi(r)$ at $r$.  That is, even without general
relativistic light deflection $\Delta\phi$ can be nonzero.  For
example, if $\psi=\pi/2$ (so that the photon is emitted 
tangentially to the radial vector), then if $r\gg M$ then 
$\Delta\phi=\pi/2$.  After computing $\Delta\phi$,
the angular location in the global coordinate system
can be calculated as before, by rotating from the temporary
equatorial plane to the global system.

\subsection{Boundary Conditions}

The initial location and direction of propagation 
of the photons depend on the run.
The default condition is that the photons
start on the surface of the star and are beamed directly outward
(we will refer to this as a ``pencil beam" specific intensity).
We will, however, consider other specific intensities, such as
one that is isotropic outwards or one that has the slightly
beamed pattern appropriate for radiation that was generated deep
in the star and that propagated outward via isotropic scattering
(see Chandrasekhar 1960, chapter 3). The individual scatters are
assumed to be locally isotropic.

Note that we make a distinction between the angular width of
the specific intensity and the angular pattern of emission on
the star. The former is what an observer standing on the star
would measure from a particular emitting point, whereas the
latter is the variation in total intensity (integrated over
local angles) as a function of position on the star.

\section{ATTENUATION OF BEAMING PATTERNS WITHOUT SCATTERING}

Before treating the effects of scattering, we first note that
a beaming pattern of high amplitude at the surface of
the star may appear to an observer at infinity to have a low
or zero amplitude. This could happen because the angular width of the
specific intensity from radiating points on the stellar surface is 
nonzero, and thus an observer at infinity sees light from
everywhere on the visible portion of the star. If the specific
intensity has a wide beaming pattern (e.g., if the pattern
is isotropic), then the intensity seen by the observer
integrates over much of the star, and radiation patterns
with large numbers of lobes are smeared out at infinity,
leading to low amplitudes of intensity modulation as the star
rotates.

In this section we derive expressions for the modulation amplitude
seen at infinity, with no scattering, under various assumptions
about the specific intensity and the angular pattern of emission
on the star. We first treat the case of straight-line photon
propagation, and show that for an isotropic specific intensity and
an odd number $n>1$ of lobes in the
stellar emission pattern the intensity seen at infinity is
constant. Thus, the relative modulation amplitude measured at
infinity can be much less than the relative amplitude measured
at the source even without light deflection.
We then consider general relativistic
light deflection. We confirm that, as demonstrated before
(e.g., Pechenick et al. 1983), light deflection has a tendency
to decrease the modulation amplitude. However, we also show
that for some harmonics of the stellar spin frequency, the 
modulation seen at infinity can actually have a {\it higher}
amplitude when light deflection is included,
compared to the modulation that would be observed if the photon
trajectories were straight. In our treatments of both the
straight and curved photon trajectories, we assume that the
emission pattern on the star is a thin equatorial belt, with
a half-thickness $h$ much less than $R$. This is intended to
model the intensity distribution expected in beat-frequency models
of kilohertz QPOs.

\subsection{Straight-Line Photon Propagation}

Assume that the
specific intensity is isotropic outwards, and that the
emission intensity at an azimuthal angle $\phi$ is
\begin{equation}
I(\phi)=I_0+\sum_{n=1}^{\infty} I_n\cos(n\phi)\; .
\end{equation}
Here $n$ gives the $n$th harmonic; $n=1$ is the fundamental,
$n=2$ is the first overtone, and so on.
The flux observed at infinity
from a short segment of the equatorial belt is proportional
to the product of the projected area of the segment (which is
proportional to the cosine of the angle $\xi$ between the line 
of sight and
the surface normal) and the emission intensity of the segment.
Let the observer be at an angle $(\theta,0)$, and
assume that the segment of interest is at the angle
$(\pi/2,\phi)$. Then $\cos\xi=\sin\theta\cos\phi$, and thus
emission is observed between $\phi=-\pi/2$ and $\phi=\pi/2$.
If the star rotates with angular frequency $\omega$, then
at a time $t$ the star is at a rotational phase $\omega t$
and the observed intensity is
\begin{equation}
\begin{array}{l}
I\propto \sin\theta\int_{-\pi/2}^{\pi/2}\left[I_0+
\sum_{n=1}^\infty I_n\cos(n(\phi+\omega t))\right]
\cos\phi\,d\phi\\
= \sin\theta\left[2I_0+I_1{\pi\over 2}\cos\omega t+
\sum_{m=1}^\infty I_{2m}{2\over{4m^2-1}}\cos 2m\omega t\right]\; .
\end{array}
\end{equation}
Hence, the modulation amplitude vanishes for
odd-numbered harmonics other than the fundamental, and decreases
as $\sim n^{-2}$ for the even harmonics. If the specific intensity
is proportional to $\sin^m\xi$, then the variable part of the
observed flux vanishes if $n+m$ is odd and $n\neq m+1$, and is
finite otherwise.
Thus, contingent on the angle dependence of the specific intensity,
either odd or even harmonics can integrate to zero.

\subsection{Curved Photon Trajectories}

When light deflection is included, the observer can see more
of the star. A larger fraction of the emission is thus
observed, and hence one might expect that the general tendency
will be for light deflection to reduce the observed amplitude
modulation. Indeed, this is the case at, e.g., the fundamental
of the rotation frequency when the specific intensity is
isotropic. However, since the constancy of the flux when
$n\neq m+1$ and $n+m$ is odd comes from the vanishing of the
integral when the limits are exactly $-\pi/2$ and $\pi/2$,
the change of these limits by light deflection means
that the amplitudes at these harmonics are actually {\em larger}
when light deflection occurs than they would be in the limit
of straight-line photon propagation.
We show this effect in Figure~1, in which we plot the rms
amplitude vs. $M/R$ for different numbers $n$ of lobes in the
emission pattern (i.e., different harmonic numbers $n$).

\begin{figure}[t]
\vskip -0.3truein
\vbox{\hskip1.5truein{
\psfig{file=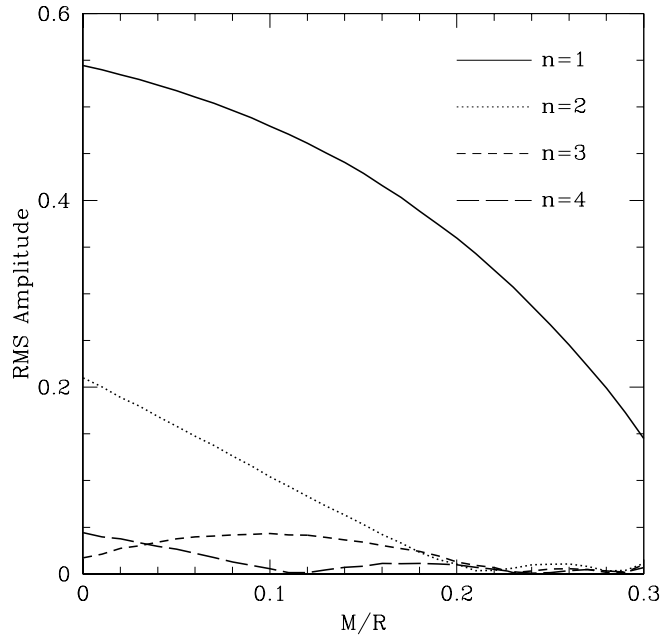,height=3.5truein,width=3.5truein}}}
\caption[fig1]{
Fractional rms amplitude versus compactness of a
neutron star, without scattering. We assume a specific intensity
at the surface that is isotropic outwards.  The solid line is for a
radiation emission pattern with one lobe (i.e., intensity
proportional to $1+\cos\phi$, where $\phi$ is the azimuthal angle
on the stellar surface).  This is therefore a pattern that would
produce a pure sinusoid at the fundamental of the spin frequency.
If there is a nonzero background and the intensity is therefore actually
proportional to $1+A\cos\phi$, where $A<1$, then the amplitudes
in this figure are multiplied by $A$.  Similarly, the dotted line
is for two lobes (i.e., the first overtone, or the second harmonic),
the short
dashed line is for three lobes (the second overtone, or third harmonic),
and the long dashed line is for
four lobes (the third overtone, or fourth harmonic).
This figure demonstrates that light deflection usually decreases,
but can increase, the fractional rms of an oscillation
as observed at infinity, compared to the case of straight-line
photon propagation, which is equivalent to the $M/R\rightarrow 0$
results in this figure.}
\end{figure}

\section{SCATTERING AROUND FINITE OBJECTS}

Shortly after low-frequency QPOs were first discovered in neutron-star
LMXBs (van der Klis et al. 1985; for a review see van der Klis 1989), 
a beat-frequency model was suggested
for them in which the frequency of the QPO is equal to the
difference between the stellar spin frequency and the frequency
of a Keplerian orbit near the main magnetospheric gas pickup radius
(Alpar \& Shaham 1985; Lamb et al. 1985; Shibazaki \& Lamb 1987). 
In such a model, one would also expect to see
a peak in the power density spectrum
at the stellar spin frequency itself, but no such
peak was observed. An early qualitative idea (Lamb et al. 1985; 
Lamb 1986), backed up by more quantitative calculations (Brainerd
\& Lamb 1987; Wang \& Schlickeiser 1987; Kylafis \& Phinney 1989),
was that whereas the beat frequency oscillation is a luminosity
oscillation and thus not easily attenuated, the brightness oscillation
expected at the stellar spin frequency is a beaming oscillation and
is thus relatively easily isotropized and attenuated by scattering
through a hot central corona.

The existence of such a scattering corona near the neutron stars
in LMXBs has been inferred by detailed comparisons of spectral
models with observations (see, e.g., Psaltis, Lamb, \& Miller 1995). 
These model fits give radial optical depths
to Thomson scattering of $\tau\sim 2\dash 10$ and scattering corona
radii of $r_c\sim 1\dash 3\times 10^6$~cm (see Miller et al.\ 1998). 
With such radii and optical
depths, the beaming oscillation at the spin frequency is expected
to be attenuated by a factor of several, which would lower the
observed amplitudes to values consistent with current upper limits.

Now that RXTE has observed a number of these sources and discovered
high-frequency QPOs, we are once again faced with the quantitative
question of why the spin frequency or other harmonics of the
kilohertz Keplerian frequency are not observed in the power spectra
from these LMXBs. Due to the high quality of RXTE data and the more
stringent upper limits it provides, we need to consider a number
of the effects that previous treatments neglected for simplicity.
We do this in this section, where we consider (1)~a beaming
pattern that originates from the surface of a finite-sized neutron
star instead of from the center of the scattering cloud, (2)~a 
corona with a number density that changes with radius, (3)~beaming
patterns for different harmonics, (4)~different specific intensity
distributions, and (5)~the effects of general relativistic
light deflection. Our
most important results are that the finite size of the neutron
star decreases the expected attenuation of a beaming oscillation
as seen at infinity even if the corona is much larger than the
star, and that for many emission geometries {\it all} overtones
(not just the even harmonics) are attenuated far more than
is the fundamental.

\subsection{Two-Stream Analysis of Expected Modulation}

To follow the treatment of Brainerd \& Lamb (1987),
consider a plane-parallel slab that goes from optical depth
$-\tau_c$ to $\tau_c$, and assume that there is a source of
intensity $F_0$ pointing forward (towards $\tau=\tau_c$) at
$\tau=\tau_R$. If we solve
this problem in the two-stream approximation, where $I_f(\tau)$
is the forward intensity at $\tau$ and $I_b(\tau)$ is the backward
intensity at $\tau$, the equations are

\begin{equation}
\begin{array}{l}
{dI_f\over{d\tau}}=-{1\over 2}I_f+{1\over 2}I_b+
F_0\delta(\tau-\tau_R)\\
{dI_b\over{d\tau}}=-{1\over 2}I_f+{1\over 2}I_b\; .
\end{array}
\end{equation}
with the boundary conditions $I_f(-\tau_c)=I_b(\tau_c)=0$.

Solving these equations, we find that 
\begin{equation}
\begin{array}{l}
I_f(\tau_c)={1\over 2}
F_0{1\over{1+\tau_c}}\left(2+\tau_c+\tau_R\right)\; , {\rm and}\\
I_b(-\tau_c)={1\over 2}F_0{1\over{1+\tau_c}}\left(
\tau_c-\tau_R\right)\\
\end{array}
\end{equation}

A measure of the asymmetry of the emission is the beaming ratio

\begin{equation}
{I_f-I_b\over{I_f+I_b}}={1+\tau_R\over{1+\tau_c}}\; .
\end{equation}
This confirms the result of Wang \& Schlickeiser (1987), who
showed that in the diffusion approximation ($\tau_c\gg 1$,
$\tau_R\gg 1$) the anisotropy of radiation depends approximately
on the ratio $\tau_R/\tau_c$.
Brainerd \& Lamb (1987) effectively considered the 
special case $\tau_R=0$.
Note that the offset of the emission from the center of the
scattering cloud multiplies the modulation
amplitude seen at infinity by the constant factor $1+\tau_R$, 
and does {\it not}
asymptote to the $\tau_R=0$ amplitude even when $\tau_c\gg \tau_R$.
The results of the numerical calculation with a finite object
are shown in Table~1, where both the star and the scattering cloud
are assumed to be spherical. In this table, $10^4$ photons
were used for each calculation, and each one started out
on the surface of the star at an optical depth $\tau_R$ from the 
center and was directed radially outward (thus, this
is a result for an initial pencil beam). The numbers are the
ratios of forward to backward intensity, defined as the ratio
of the emergent intensity through the northern hemisphere
($\cos\theta>0$) to the emergent intensity through the southern
hemisphere.  The numbers in
parentheses are for no central star but a photon that starts
at radius $\tau_R$; this comparison indicates the importance of
a hard surface vs. the importance of an initial offset from
the center.

This table shows that the presence of a central star increases the
beaming ratio significantly, when compared to the beaming ratio
produced without a finite central star, both by the offset of the
initial emission from the center of the scattering cloud and by
the presence of a hard surface. The analytical estimate
for the beaming ratio matches the numerical
results well when there is no star but the photon starts at an
optical depth $\tau_R$.

\subsection{Radius-Dependent Number Density}

Now consider what happens when the electron number density is not 
constant with radius. We explore three different density
profiles. First, a density that goes as $n\sim (r_0^2-r^2)$,
where $r_0$ is the radius of the scattering region.
Second, a density that is $n=2n_0$ for $0\leq r\!<\!r_0/2$ and
$n=n_0$ for $r_0/2\leq r\leq r_0$. Third, a density that
is $n=n_0$ for $0\leq r\!<\!r_0/2$ and $n=2n_0$ for $r_0/2\leq 
r\!<\!r_0$.
In Table~2 we compare the beaming ratios for these
three different density profiles (``Quad", ``2$\rightarrow$1",
and ``1$\rightarrow$2", respectively), for $\tau_R=0$ and the same total
optical depth $\tau_{\rm total}$ from the center to $r_0$.  In each
case we assume a pencil beam specific intensity originating from
the center of the scattering cloud.
This table shows that, compared to a uniform-density scattering
cloud, a cloud with an edge
concentration yields a higher anisotropy as seen at infinity.
This is because the lower the number density is close in, the farther
photons can travel before scattering.  For a pencil-beam specific
intensity this is effectively similar to having the source of the
radiation be offset from the center by a relatively large distance,
implying that the anisotropy of the emergent radiation is larger
than it would have been for uniform density.  Conversely, a cloud
with a central concentration yields a lower anisotropy at infinity
than if the cloud had uniform density.

\subsection{Attenuation of Overtones}

Using the same treatment as Brainerd \& Lamb (1987), 
we find that as $\tau_c\rightarrow\infty$
the scattering Green's function approaches

\begin{equation}
G(\theta,\phi,\tau_c)={1\over{4\pi}}\left[1+2\left(1+\tau_R\over{
1+\tau_c}\right)\cos\theta\right]\; ,
\end{equation}
where $\theta$ is the angular distance between the observer and the
point of emission and there is no dependence on the longitude $\phi$
of the observer.  We wish to determine the intensity seen at infinity
if the emission pattern on the surface has some number of lobes
(equivalently, for some harmonic).
Consider first the lobe structure used by Brainerd \& Lamb,
which is that the intensity on the surface is proportional to
$1+\cos(n\theta)$. This intensity pattern is appropriate for an
emission that has a favored axis, such as a centered dipole.

If the center of the emission is at an angle $(\theta^\prime,0)$ and the
observer is at an angle $(\theta,\phi)$, then the
cosine of the angle between them is just $\cos\psi=
\sin\theta\sin\theta^\prime\cos\phi+\cos\theta\cos\theta^\prime$.
The intensity seen at $(\theta,\phi)$ is

\begin{equation}
I(\theta,\phi,\tau_c)=\int_0^{2\pi}\int_0^\pi 
G(\psi,\phi,\tau_c)
I(\theta^\prime)\sin\theta^\prime d\theta^\prime d\phi^\prime\; ,
\end{equation}
where $I(\theta^\prime)$ is the intensity at the stellar surface at the
colatitude $\theta^\prime$.  The $\phi$ term will therefore always 
integrate to zero, regardless of the number of lobes. The intensity is then

\begin{equation}
\begin{array}{ll}
I(\theta,\phi,\tau_c)&\sim {1\over 2}\int_0^\pi \left[1+
2\left(1+\tau_R\over{1+\tau_c}\right)\cos\theta\cos\theta^\prime\right]
(1+\cos n\theta^\prime)\sin\theta^\prime d\theta^\prime\\
&=\left\{
\begin{array}{ll}
{1\over{4\pi}}\left[1+\left(2\over{4-n^2}\right)\left(1+\tau_R\over{
1+\tau_c}\right)\cos\theta\right]I_0\; ,& n{\rm \ odd} \\
{1\over{4\pi}}I_0\; ,& n{\rm \ even}
\end{array}
\right.
\end{array}
\end{equation}
in the limit $\tau_c\gg 1$.  Here we have assumed $I(\theta^\prime)=
I_0(1+\cos n\theta^\prime)$.  Note that a fraction $\sim e^{-\tau_c}$ of
the photons will escape directly, and hence for finite $\tau_c$
the amplitude at even harmonics is nonzero but small.
This extends the result of Brainerd \& Lamb (1987) to photons
offset from the center of the scattering cloud:
for emission symmetric around an axis, even-lobed patterns
are attenuated much more rapidly than are odd-lobed patterns,
and odd-lobed patterns with $n>1$ are more rapidly attenuated
than is the $n=1$ pattern.

Consider now a pattern that is symmetric about the equatorial
plane but is not axisymmetric.  This is the pattern of interest
for any rotationally or orbitally modulated emission, such as
the emission believed to produce the higher-frequency QPO peaks
observed during persistent emission from neutron-star LMXBs, 
which is symmetric about the rotational equator but is otherwise
arbitrary. Specifically, consider an
emission pattern $I_0(\theta^\prime,\phi^\prime)\sim 
[1+T(\theta^\prime)\cos n\phi^\prime]$, where $T(\theta^\prime)
=T(\pi-\theta^\prime)$. In this case, the intensity seen at
infinity in the direction $(\theta,\phi)$ is

\begin{equation}
I(\theta,\phi,\tau_c)\sim\int_0^\pi \int_0^{2\pi} G(\psi,\phi,\tau_c)
(1+T(\theta^\prime)\cos n\phi^\prime)\sin\theta^\prime d\theta^\prime
d\phi^\prime\; ,
\end{equation}
where $\psi$ is the angle between $(\theta,\phi)$ and
$(\theta^\prime,\phi^\prime)$:
\begin{equation}
\cos\psi=\sin\theta^\prime\sin\theta\sin\phi^\prime\sin\phi+
\sin\theta^\prime\sin\theta\cos\phi^\prime\cos\phi+
\cos\theta^\prime\cos\theta\; .
\end{equation}

The $\phi-$dependent terms in this integral are proportional
to either $\int_0^{2\pi}\sin\phi^\prime\cos n\phi^\prime 
d\phi^\prime$ or $\int_0^{2\pi}\cos\phi^\prime\cos n\phi^\prime
d\phi^\prime$. However, note that for $n\neq 1$

\begin{equation}
\begin{array}{l}
\int_0^{2\pi}\sin\phi^\prime\cos n\phi^\prime d\phi^\prime=
\left[{\cos(n-1)\phi^\prime\over{2(n-1)}}-{\cos(n+1)\phi^\prime
\over{2(n+1)}}\right]_0^{2\pi}=0\; ,\ {\rm and} \\
\int_0^{2\pi}\cos\phi^\prime\cos n\phi^\prime d\phi^\prime=
\left[{\sin(n-1)\phi^\prime\over{2(n-1)}}+{\sin(n+1)\phi^\prime
\over{2(n+1)}}\right]_0^{2\pi}=0\; .
\end{array}
\end{equation}
Note also that the $\cos\theta^\prime T(\theta^\prime)$ term
integrates to zero, due to the symmetry of $T(\theta^\prime)$.
Therefore, in the diffusion limit ($\tau_c\gg 1$), 
$I(\theta,\phi,\tau_c)\rightarrow 0$ for $n>1$.

This result means that for {\it any} emission pattern that is
symmetric about the rotational
equator, {\it all} overtones are attenuated extremely rapidly,
not just the even harmonics. In particular, note that it
is not necessary to have the same $\phi^\prime$ dependence
at all latitudes; an arbitrary emission pattern symmetric
about the equator can be built up using pairs of rings of
the form $\delta(\theta^\prime-\theta_0)H(\phi^\prime)+
\delta(\theta^\prime-[\pi-\theta_0])H(\phi^\prime)$, where
$H(\phi^\prime)$ is some function of $\phi^\prime$ ($H$
can, therefore, be Fourier decomposed into terms proportional
to $\cos n\phi^\prime$). This is a strong reason why, even
if the fundamental is strong at the sonic-point Keplerian
frequency, we do not expect to see significant peaks in the
power spectrum at overtones of $\nu_{\rm Ks}$.

Figure~2 shows the fractional rms amplitude of different harmonics
as a function of the optical depth of the scattering cloud.
Here we assume a pencil-beam
specific intensity and straight-line photon propagation, and we
assume that the central
object has a radius of $\tau_R=1$ optical depths.  Therefore,
$\tau_c=1$ would mean that there is no scattering cloud; $\tau_c=2$
would mean that the radial optical depth from the stellar surface to
the edge of the cloud is $\tau_c-\tau_R=1$, and so on.
As in Figure~1, $n$ is the number of lobes in the radiation
emission pattern from the surface; thus $n=1$ is the fundamental
of the spin frequency, $n=2$ is the first overtone, and so on.
For harmonic number $n$ we assume an emission intensity proportional 
to $1+\cos n\phi$; for an intensity actually
proportional to $1+A\cos n\phi$, with $A<1$, the magnitude of the
amplitude must be multiplied by $A$.  This figure shows that all
overtones of the fundamental oscillation frequency are attenuated
rapidly by scattering.

\begin{figure*}[thb]
\begin{minipage}[t]{3.2truein}
\mbox{}\\
\psfig{file=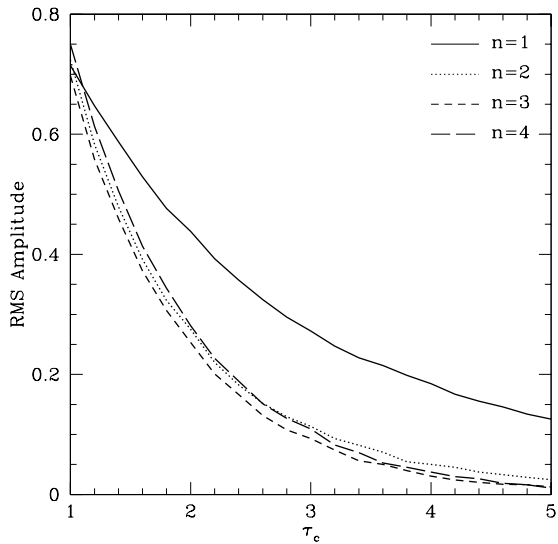,height=3.0truein,width=3.0truein}
\end{minipage}
\hfill
\begin{minipage}[t]{3.2truein}
\caption[fig2]{
Fractional rms amplitude for harmonic numbers $n$ versus the optical depth
of a surrounding scattering cloud, assuming a pencil-beam
specific intensity, straight-line photon propagation, and $\tau_R=1$.
This figure shows that for rotationally modulated
beaming oscillations the amplitudes at {\it all} overtones of the
fundamental oscillation frequency are decreased very rapidly
by propagation through a scattering cloud. This is an important
reason why, to date, no overtones of Keplerian orbital frequencies
near neutron stars have been observed from low-mass X-ray
binaries.}
\end{minipage}
\end{figure*}

\subsection{The Angular Dependence of the Specific Intensity}

Heretofore we have considered only a pencil beam type of
specific intensity. In reality, the angular dependence of the
specific intensity is likely to differ from a pencil beam.  In
Figure~3 we show the rms amplitude vs. optical depth for
a pencil beam, for the beamed pattern appropriate 
for radiation generated at great depth (see Chandrasekhar 1960,
chapter 3), and for an
isotropic beam. These different specific intensity distributions
have different uses. A pencil beam may be considered to set an
upper limit on the anisotropy, but it is unlikely to be of
direct significance in the physical situations considered here.
The beamed pattern appropriate for
radiation generated at great depth that propagates to the
surface via isotropic scattering is likely
to represent well the emergent radiation pattern from an
X-ray burst. An isotropic specific intensity may be a good
model for the radiation produced by accretion, since the
energy is likely to be released in a shallow layer. From this
figure, we see that, as expected, the more beamed the pattern
the greater the rms amplitude of variation for a given
surface intensity distribution. We also see that the
Chandrasekhar-type specific intensity gives a modulation
amplitude closer to an isotropic beam than to a pencil beam.

\begin{figure*}[thb]
\begin{minipage}[t]{3.2truein}
\mbox{}\\
\psfig{file=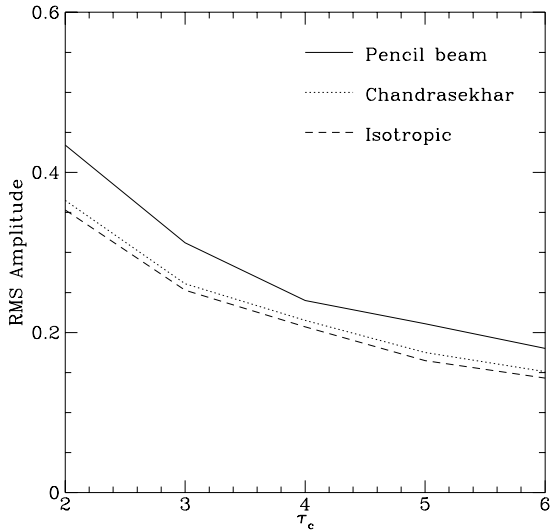,height=3.0truein,width=3.0truein}
\end{minipage}
\hfill
\begin{minipage}[t]{3.2truein}
\caption[fig3]{
Fractional rms amplitude versus optical depth for
different beaming patterns at the surface.  We assume straight-line
photon propagation.  Here $\tau_R=1$,
and the amplitudes are for a single lobe (i.e., the fundamental
of the oscillation frequency), with emission intensity proportional
to $1+\cos\phi$.  As expected, the
more beamed the intensity, the higher the oscillation amplitude.
The beaming pattern derived by Chandrasekhar, which is the
emergent specific intensity for radiation generated deep below
the surface that propagates upward by isotropic scattering, gives
amplitudes very close to the isotropic case.}
\end{minipage}
\end{figure*}

\subsection{The Effects of Light Deflection}

As we described in \S~2, the code has the capability to follow
curved photon trajectories. We show the effects of light deflection
in a scattering cloud in Figure~4 and Figure~5. Figure~4 plots
the rms amplitude vs. $M/R$ for different numbers of nodes for
a pencil beam with $\tau_R=1$, $\tau_c=5$. Figure~5 compares the
amplitude vs. $M/R$ for $\tau_R=1$, $\tau_c=5$ for a pencil beam
(solid line),
an isotropic beam (dotted line), and a pencil beam which undergoes no light
deflection after it escapes from the scattering cloud (dashed line). 
The curves in Figure~5, especially those for the pencil beam with
and without deflection after escape, demonstrate that there are actually
two effects on the amplitude that compete with each other.  The deflection
of light after escape spreads out the beam, and hence decreases the
amplitude.  The deflection of light between scatters, however, can have the
opposite effect.  For a given distance traveled after one scatter, the
subsequent scatter is angularly closer to the first because the photon
travels in a curved trajectory.  This tends to increase the number of
scatters, because the coordinate distance traveled is less between 
scatterings.  However, for a fixed number of scatters the angular distance 
traveled is smaller, and hence the effective isotropization is diminished.
The overall effect is always that the modulation amplitude decreases with
increasing compactness $M/R$, but Figure~5 shows that
for $M/R>\sim{1\over 6}$ the amplitude would increase with increasing
$M/R$ if there were no deflection after escape.

\begin{figure*}[thb]
\begin{minipage}[t]{3.2truein}
\mbox{}\\
\psfig{file=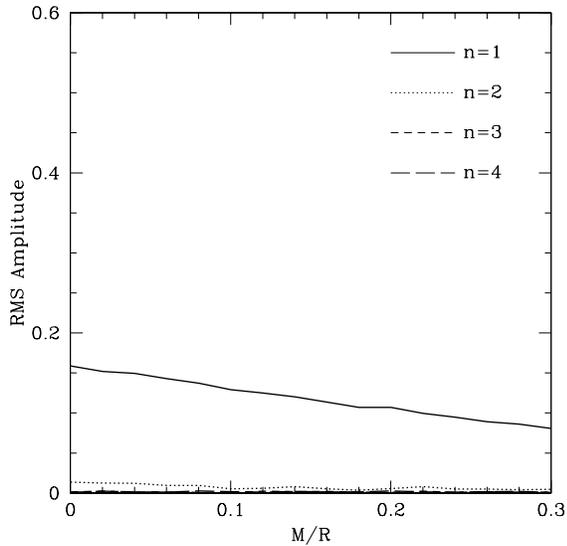,height=3.0truein,width=3.0truein}
\end{minipage}
\hfill
\begin{minipage}[t]{3.2truein}
\caption[fig4]{
Fractional rms amplitudes versus compactness of
a neutron star for different numbers $n$ of lobes (equivalently,
for different harmonic numbers $n$), where as before we assume
a total intensity proportional to $1+\cos n\phi$.  We assume
straight-line photon propagation.  Here the initial
beaming pattern is a pencil beam, and it is clear that the amplitude
at any overtone is small. The stellar radius, in optical depths,
is $\tau_R=1$ and the radius of the scattering cloud in optical
depths is $\tau_c=5$.}
\end{minipage}
\end{figure*}

\begin{figure*}[thb]
\begin{minipage}[t]{3.2truein}
\mbox{}\\
\psfig{file=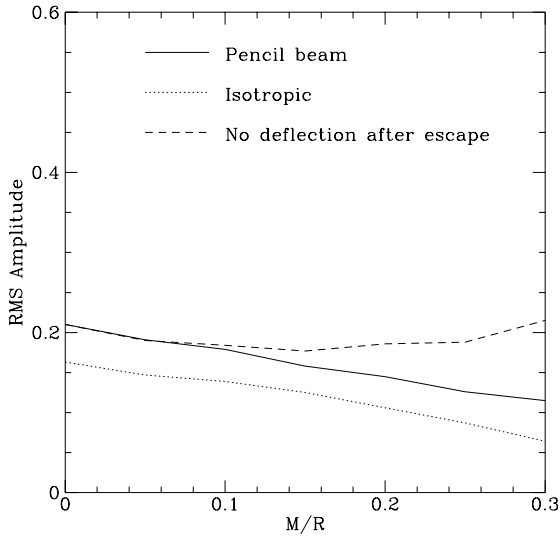,height=3.0truein,width=3.0truein}
\end{minipage}
\hfill
\begin{minipage}[t]{3.2truein}
\caption[fig5]{
Fractional rms amplitudes versus compactness of a
neutron star for (a)~a pencil beam, (b)~an isotropic beam, and
(c)~an artificial case in which the photons travel in straight
lines after escape from the scattering cloud (for this curve
we assume a pencil beam). As in Figure~4,
$\tau_R=1$ and $\tau_c=5$.  Here we assume a single lobe (i.e.,
the fundamental of the oscillation frequency), with intensity
proportional to $1+\cos\phi$.}
\end{minipage}
\end{figure*}

\section{MODULATION AMPLITUDE INSIDE A SCATTERING CLOUD}

Another important question is how great a modulation amplitude
is to be expected {\it inside} a scattering cloud. For example,
in the sonic-point model (Miller et al.\ 1998) 
the lower-frequency QPO in a pair
is generated by the interaction of radiation at the surface with
clumps of matter in the disk. The radiation is modulated at
the stellar spin frequency, and thus the mass accretion rate
from the clump is modulated at the difference between the
sonic-point Keplerian frequency and the stellar spin frequency.
From no source with kilohertz QPOs has there been a strong peak
at the stellar spin frequency detected in the power density spectrum, and
hence it is important to determine if it is possible that the modulation
at the spin frequency is strong near the star, where the beat
frequency is generated, yet weak enough at infinity that no
peak in the power density spectrum is evident.  It is
not easy to compute the modulation in mass accretion rate that
results from a given modulation amplitude in the radiation force
or any of the components of the radiation stress-energy tensor,
because the effects of radiation drag can be nonlinear. For example,
since extra radiation drag increases the radial velocity of gas
and thus decreases the optical depth from the stellar surface 
(``radiation-induced transparency"; see Miller \& Lamb 1996;
Miller et al.\ 1998), 
a small increase in radiation drag can in principle lead to
a large increase in the accretion rate. Nonetheless, as a proof of
principle we can calculate the radiation energy density, to
determine if it is modulated significantly near the star.

Figure~6 shows the result of this calculation. Here we have chosen
$\tau_R=1$ and $\tau_c=5$, for illustrative purposes, and we assume
a central star of radius $R=5M$. The solid
line shows the fractional rms amplitude of the modulation in the
radiation energy density,
and the horizontal dotted line shows the fractional rms amplitude in
the energy density as observed at infinity. It is clear from this
figure that the modulation amplitude can be much higher near
the star than at infinity. For example, the rms amplitude
is more than ten times as high
at the surface of the star as it is at infinity. This confirms that a
brightness oscillation can have strong effects near the star,
but be relatively weak at infinity.

\begin{figure*}[thb]
\begin{minipage}[t]{3.2truein}
\mbox{}\\
\psfig{file=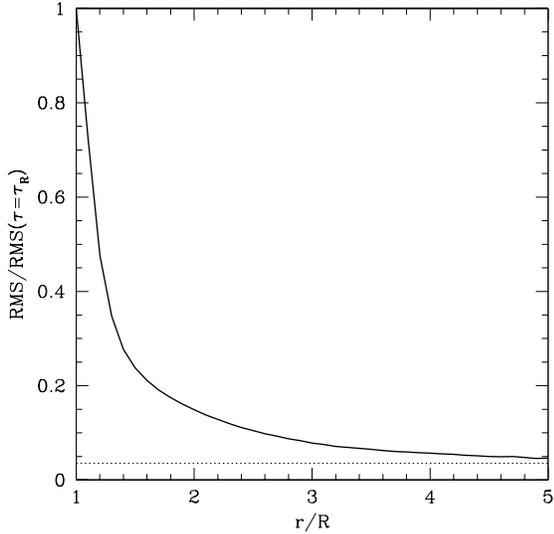,height=3.0truein,width=3.0truein}
\end{minipage}
\hfill
\begin{minipage}[t]{3.2truein}
\caption[fig6]{
Root mean square amplitude of modulation of the energy density
(the ${\hat t}{\hat t}$ component of the stress-energy tensor)
divided by the rms amplitude at the stellar surface.
Here $\tau_R=1$ and $\tau_c=5$, and the stellar radius is $R=5M$
(we therefore include light deflection in this calculation).  For
this figure we assume pencil-beam emission from a single spot on
the star.  The dashed horizontal line
is the rms amplitude (divided by the amplitude at $r=R$) observed
at infinity.  The high ratio of amplitude near the star to amplitude
at infinity shows that it is possible to have
a brightness oscillation of relatively large amplitude near the star that
is weak or undetectable far from the star.}
\end{minipage}
\end{figure*}

\section{DISCUSSION AND CONCLUSIONS}

The amplitude of beaming oscillations as seen at infinity depends
on a number of variables, including the radiation pattern on the
surface of the star, the angular width of the specific intensity,
the size of the star, and the size, optical depth, and density
profile of the
scattering cloud. Some of the trends evident from this paper are
(1)~when there is a scattering cloud, the observed amplitudes of all
overtones are much less than the observed amplitude of the fundamental,
(2)~a brightness oscillation that is weak or undetectable with current
instruments at infinity
can nonetheless have a significant amplitude near the star, and
(3)~the presence of a finite-sized star in the center of a
scattering cloud decreases the attenuation of beaming oscillations
compared to the attenuation expected when the photons are emitted
at the center of the cloud.

These results are particularly useful in the study of the kilohertz
QPOs from neutron-star LMXBs. Spectral models (see
Lamb 1989; Miller \& Lamb 1992; Psaltis et al. 1995) suggest that
many of the neutron stars in LMXBs are surrounded by hot central coronae
with radii of $1\dash 2\times 10^6$~cm and
optical depths $\tau\sim 3\dash 10$. The observed amplitudes of kilohertz
QPOs, combined with a detailed model of them, can be used to place
further restrictions on the radius and optical depth of the central
corona, and possibly even on the compactness of the neutron star.

In conclusion, the results presented here on attenuation of
beaming oscillations explain several features of the kilohertz
QPOs, including why no overtones of a beaming oscillation have
been detected and why the beaming oscillation at the stellar
spin frequency can be strong enough near the star to generate
a beat-frequency QPO (see Miller et al.\ 1998), 
yet too weak to detect at infinity.
The attenuation factors at frequencies such as the Keplerian QPO 
frequency, the spin frequency, and the overtones and sidebands of these
fundamental frequencies depend on the radius and optical depth
of the scattering cloud and on the redshift at the surface of
the neutron star, and hence detections of QPOs at these frequencies,
or strong upper limits on their amplitudes, provide a valuable source
of information about the conditions near neutron stars.

\acknowledgements

It is a pleasure to thank Fred Lamb and Dimitrios Psaltis
for discussions about the optical depths and radii of coronae
in LMXBs, and Don Lamb and Carlo Graziani for their comments
on early versions of the ideas contained in this paper. This
work was supported in part by NASA grant NAG 5-2868,
by NASA ATP grant number NRA-98-03-ATP-028, and through the GRO 
Fellowship Program by NASA grant NAS 5-2687.

\newpage

\begin{table*}[t]
\centering
\begin{tabular}{ccccccc}\hline
\multicolumn{7}{c}{\bfseries Table 1: Beaming Ratios With a
Central Object}\\
\hline
$\tau_c$&$\tau_R=0$&$\tau_R=1$&$\tau_R=2$&$\tau_R=3$&$\tau_R=4$&$\tau_R=5$\\
1&0.465&---&---&---&---&---\\
2&0.301&0.697(0.605)&---&---&---&---\\
3&0.246&0.528(0.459)&0.810(0.712)&---&---&---\\
4&0.199&0.448(0.364)&0.678(0.566)&0.877(0.774)&---&---\\
5&0.174&0.380(0.311)&0.601(0.486)&0.769(0.643)&0.906(0.814)&---\\
10&0.094&0.218(0.183)&0.337(0.273)&0.453(0.366)&0.577(0.461)&0.699(0.556)\\
20&0.056&0.121(0.085)&0.184(0.145)&0.254(0.205)&0.332(0.234)&0.391(0.299)\\
\hline
\end{tabular}
\end{table*}

\begin{table*}[thb]
\centering
\begin{tabular}{ccccc}\hline
\multicolumn{5}{c}{\bfseries Table 2: Beaming Ratios With}\\
\multicolumn{5}{c}{\bfseries Different Density Dependences}\\
\hline
$\tau_{\rm total}$&Constant&Quad&2$\rightarrow$1&1$\rightarrow$2\\
1&0.465&0.453&0.462&0.487\\
2&0.301&0.273&0.268&0.334\\
3&0.246&0.206&0.206&0.286\\
4&0.199&0.173&0.152&0.224\\
5&0.174&0.127&0.134&0.195\\
10&0.094&0.082&0.070&0.109\\
20&0.056&0.041&0.033&0.055\\
\hline
\end{tabular}
\end{table*}

\end{document}